# Structured Light Projection-based single-shot transport of intensity phase imaging for quantitative diagnostics of high-speed flows

Biswajit Medhi[1] and K.P.J. Reddy

Indian Institute of Science, Bangalore-12, India

**Contributions:**

1. A method for quantitative density estimation in high speed flows is presented that uses a Structured light beam to interrogate the flow field, where the images of the flow are captured through customized camera optics designed for this purpose.

2. A technique is proposed that connects the spatial distribution of the geometrically projected structured light beam to its axial intensity derivative, $\partial I/\partial z$, thus, the need for small separation distance ($\partial z$) for estimating $\partial I/\partial z$ (Finite difference approximated) is avoided and hence a low $f/\#$ imaging system is no longer required in experiments while capturing images.

3. Moreover, the stringent requirement of a sensitive camera to see minute intensity variations at different planes is relaxed as $\partial I/\partial z$ is no longer estimated through absolute intensities.

4. The transport of intensity equation (TIE) is used to estimate the phase of the interrogating beam quantitatively followed by density recovery through iterative phase tomography, where the experimentally derived intensity-derivative is used as an input.

5. The construction of the proposed technique keeps the region of interest focused while maintaining a tunable sensitivity to see refractive gradients targeting a wide range of flows.

**Abstract:**

A method for quantitative estimation of density in a high-speed flow field is presented, which uses a structured-light beam probe to interrogate the region of interest (ROI). Wavefront distortion suffered by the interrogating beam as it passes through the shock-induced flow field is investigated. Dedicate camera optics is used to measure the cross-sectional intensities of the exiting wavefront at two different planes, simultaneously. A technique is proposed that uses the lateral displacement of the structured light pattern on both the planes to estimate the axial intensity derivative $\partial I/\partial z$, which is the input for the transport-of intensity equation (TIE) for phase estimation. By doing this, the need for small defocused distance, $\partial z$ is avoided in experiments and hence the demand of a low $f/\#$ imaging system is alleviated, and now larger $\partial z$ is admissible as the intensity derivative is not estimated through the conventional finite difference (FD) approximation. The recovered phase from TIE is used as an input to phase tomography algorithm to obtain the refractive index followed by density distributions in the flow field quantitatively. The proposed technique is verified through experiments conducted in hypersonic shock tunnel for flow Mach No. of 8.5. Estimated cross-sectional densities of the flow field around the aerodynamic test model are presented and compared with the numerically estimated values, which shows good agreement.

**Introduction:**

Optical diagnostic techniques are preferred for investigating flow fields as they can provide inertia-free measurements of a large region of interest (ROI) non-intrusively. Their applications are found in different areas of fluid dynamic studies, which provide qualitative as well quantitative information of the flow field [**1-24**]. Laboratory investigations of such flow fields are usually conducted in shock tunnel and wind tunnel facilities [6-7], where flow field parameters such as velocity, pressure, density are usually studied and measured non-intrusively through optical techniques within the ROI. Optical methods such as Particle Image Velocimetry (*PIV* ) and Doppler Velocimetry [1-5] are used to estimate the velocity distribution of low Mach no. (M<0.5) flows quantitatively. However, at higher velocities, the excessive pressure gradients in the flow field induce density gradients, hence the refractive index, $n(r)$ of the medium changes and thus optical techniques that are sensitive to refraction of the medium are used [1-24]. Experimental techniques such as schlieren, shadowgraph, background oriented schlieren (BOS), shadow casting and holography work on this fundamental principle, where the interrogating beam passing through the flow field gets distorted. The distortion of the interrogating beam coming out of the flow field is seen as intensity gradients in shadowgraph and schlieren techniques [8- 10] and as a displacement of the random-mask pattern in BOS [9-17] and shadow casting techniques [18-21]. Similarly, the interferometry encodes this gradient information in its fringe pattern as its local fringe-deformation [10-11]. The measured information by all these techniques could be processed further to retrieve the density distribution of the flow field quantitatively. In a recent study, it was shown that the intensity images captured along the optical axis at multiple planes could also be used in estimating density distribution while diagnosing hypersonic flows [22].

To the paraphernalia of flow analysis, another light-based flow diagnostic tool is proposed that uses structured light to interrogate the ROI such that quantitative information of the flow field is obtained. This is essentially an optical tomography system that collects data from wavefront trans-illuminating the flow where measurement is desired, without placing any pattern or knife-edge in the optical path. A relation is proposed that connects the spatial distortion of the geometrically projected structured light pattern to the axial intensity derivative, $\partial I/\partial z$ of the propagating beam (along the z-axis) and hence the transport of intensity equation (TIE) [23] is used to estimate the phase (see Section 2). Methods for phase estimation that rely on intensity measurements require either two measurements (TIE based) [23] separated by a very small defocused distance (such that FD approximation of intensity derivative is satisfied ) or take multiple measurements [22] to recover the phase. Experimental systems that have high $f/\#$ mirrors, has a large depth of field and cannot acquire images that can directly be fed to TIE. Here, the proposed relation that uses structured light illumination has the edge over the conventional approach, where almost any defocused distance, $\partial z$ can be utilized to estimate $\partial I/\partial z$ as the FD approximation is no longer used. Moreover, the proposed technique has the following additional features: 1) it can keep the ROI focused such that well-resolved boundary information is retrieved, 2) its sensitivity can be tuned accordingly to visualize a wide range of flow-fields, 3) only one camera is used, that eases the experimentation compared to the cumbersome multi-camera [22] approach to record dynamic events, 4) TIE is solved for images that are separated by a large distance, captured by a high $f/\#$ imaging optics.

The rest of the paper is organized as follows: Section 2 provides the theoretical background of the proposed technique and also explains the procedure to recover density through tomography. Section 3 elaborates experimental facility and the results-discussions are in Section 4. Section 5 covers the concluding remarks of the presented work.

## 2. Theoretical background:

The topic of phase estimation while observing transparent objects are perused deliberately by several researchers which are based on geometrical- and wave optics models. Geometrical optic techniques rely on relative deformation of markers (pattern: random dot, grid) when the PO is present in the optical path [18-21], from which slope of the deformed wavefront is estimated followed by wavefront recovery. However, wave optics models [22-27] are free from such positional dependencies of marker and estimate the phase straightway by using the intensity images captured along the optical axis [23]. Two standard approaches are used to estimate phase from intensities: first, analytical approach that uses logarithmic Hilbert transform [25] and the second is the conservation of energy (intensity) transfer along the optical axis. Energy conservation technique leads to two more approaches, they are either a direct solution of transport of intensity equation (TIE) derived from the Helmholtz equation under paraxial approximation [23] or iterative predictor-corrector schemes such as Gerchberg-Saxton algorithm [26] or Fienup's method (derived from the G–S algorithm [27],). The first approach, which is based on TIE is adored due to its simple experimental requirements.

The performance of TIE based imaging relies on the accuracy of approximated intensity derivative and thus it is suitable for imaging systems which has a shallow depth of focus, such as microscopy imaging [28]. Here, measured intensities ($I_1, I_2$) in two planes separated by a small defocused distance ($\partial z$) are used to estimate intensity derivative, $\partial I / \partial z (= (I_2 - I_1)/\partial z)$ (FD approximation). However, the estimation of $\partial I / \partial z$ become challenging for imaging systems that have a larger depth of field (hence $\partial z$ is large) containing high f/# objectives, as in the case of the flow visualization experiment presented in this article. Several approaches have been developed to address the phase estimation problem for a large DOF imaging system. In which, approaches that incorporate multiple intensities captured at different defocused distances are used either to estimate $\partial I / \partial z$ by some mean [22] for solving TIE or to use in the iterative predictor-corrector scheme. However, the main shortcoming of such imaging is the need for multiple images (three or more images) that become critical for dynamic events. At the same time, intensity measurements at a minimum two number of planes are necessary to estimate phase while pursuing intensity-based phase imaging. In this paradigm, a structured illumination based transport of imaging that uses modular camera optics (MCO), designed for this purpose could address this issue. In which, the MCO captures two images of the ROI at the same time that are focused at different planes along the optical axis, while structured illumination helps in getting the approximated intensity derivative, $\partial I / \partial z$ from relative spatial deformation of structured light in each plane.

The proposed technique circumvents the need for small $\partial z$, which was necessary to accurately estimate $\partial I / \partial z$ (FD approximated) from two images. Moreover, the use of structured illumination provides the flexibility to set almost any $\partial z$, enabling a variable sensitivity to detect a wide range of refraction gradients (and hence the density) of the flow field. In what follows in the current section, the details of the proposed scheme is provided along with a brief description of TIE for completeness. Other than this a brief explanation of the phase tomography scheme is also elaborated which is used to estimate the refractive index (hence density) of the flow field.

## 2.1 Transport of Intensity Equation and estimation of intensity derivative- conventional approach:

As stated earlier, the TIE is the paraxial approximation to the wave equation that connects the axial transport of intensity ($\frac{\partial I}{\partial z}$) to the phase ($\varphi(x,y)$) of the wavefront orthogonal to the optical axis of the imaging system, i.e., the $(x,y)$ plane.

The TIE is [23]

$$-k\frac{\partial I}{\partial z}=\nabla_{\perp}\cdot(I(x,y)\nabla_{\perp}\varphi(x,y)) \tag{1}$$

Here, $k$ is the modulus of the light propagation vector, the operator $\nabla_{\perp}=(i\frac{\partial}{\partial x}+j\frac{\partial}{\partial y})$ and $I(x,y)=I(x,y,z_0)$ is the intensity evaluated at a selected plane perpendicular to the optical axis, say, at $z=z_0$.

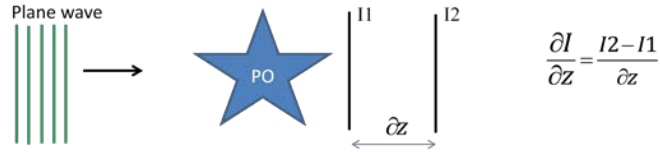

Fig. 1. Schematic of transport of intensity equation (TIE) based phase estimation technique

The phase object is interrogated through a plane wave and the exiting light wave is captured, usually with the help of a digital camera. Captured images, $I_1$ and $I_2$ in two planes are used to estimate FD approximated intensity derivative, $\frac{\partial I}{\partial z}=\frac{I2-I1}{\partial z}$, considering $\partial z\rightarrow 0$.

For a pure phase object, $I(x,y)$ is constant, and the Eq.(1) becomes $\nabla_{\perp}^{2}\varphi(x,y)=-\frac{k}{I(x,y)}\frac{\partial I}{\partial z}$; a Poisson equation. However, when $I(x,y)\neq const.$, the solution is obtained through the substitution $\nabla_{\perp}\psi(x,y)=I(x,y)\nabla_{\perp}\varphi(x,y)$, which gives another Poisson equation $\nabla_{\perp}^{2}\psi(x,y)=-\frac{k}{I}\frac{\partial I}{\partial z}$ for the unknown $\psi(x,y)$. Recovered $\psi(x,y)$ is fed back to the substitution equation to recover $\varphi(x,y)$. Spectral techniques (such as DCT, FFT etc.) are usually preferred for solving the Poisson equation which is computationally efficient. Here, DCT based technique that incorporates Neumann boundary condition is used for estimating solutions. The details of such techniques can be found in refs [7, 29].

## 2.2 Estimation of intensity derivative $\frac{\partial I}{\partial z}$ from structured illumination and solution of TIE

Estimation of $\frac{\partial I}{\partial z}$ is pursued before by several research groups for solving TIE. In which, schemes based on polynomial interpolation [30], chromatic aberration introduced by imaging lens [31], coded-aperture based focal stack imaging [32] and matrix-based interpretation are noticeable. In this paper, a new approach is proposed to estimate $\dfrac{\partial I}{\partial z}$, which is based on measured beam deflection of the interrogating beam at different planes in the optical axis. The technique is as follows:

The associated phase of a wavefront, $\qquad \varphi = k\displaystyle\int_l n\,ds$ $\qquad\qquad\qquad$ -(2)

here, $k$ - wave number, $n-$ refractive index of the media.

Now, rewriting Eq.(1) with the help of Eq.(2), and rearranging the parameters, we have

$$-k\frac{\partial I}{\partial z} = \nabla_\perp \cdot I(x,y)\left[ k\int_l \nabla_\perp n(x,y)\,ds \right] \qquad\qquad\text{-(3)}$$

The terms within the bracket on the right-hand side of Eq.(3) is an experimentally measured quantity to estimate intensity derivative, $\dfrac{\partial I}{\partial z}$. In this case, the line integral of the partial derivative of refraction i.e.

$\displaystyle\int_l \nabla_\perp n\,ds = \int_l (\hat{i}\frac{\partial n}{\partial x} + \hat{j}\frac{\partial n}{\partial y})\,ds$ gives the beam deflection angle of the interrogating beam along with the orthogonal directions while passing through the three-dimensional refractive medium (see Fig.2).

Beam deflections $\displaystyle\int_l (\hat{i}\frac{\partial n}{\partial x})ds$ and $\displaystyle\int_l (\hat{j}\frac{\partial n}{\partial y})ds$ along X and Y directions can be estimated through an experimental arrangement as shown in Fig.2, where the spatial distribution of deflected beam angles, $\boldsymbol{\theta}$ of the projected beam, can be estimated through:

$$\theta = \frac{dy_1}{D} = \frac{dy_2}{D+d} = \frac{dy}{d} = \frac{dy_2 - dy_1}{d}\ldots\ldots\ldots(4)$$

Here, the PO is interrogated through a structured illumination, which gets deflected when it sees a refractive gradient in its path that is monitored at planes ( *Plane I* , *Plane II* ). It is worth mentioning that the deflection angles could also be measured by using the information captured in any one of the planes (say *Plane I*, $^{dy_1}/_{D}$ ). However, the accuracy of measurements is compromised due to inaccurate $D$ , for an object that has an unknown spatial extension ( $d_{obj}$ ) along the optical axis, which has been addressed by incorporating one more measurement that provides $d$ , a user-defined parameter during experiments.

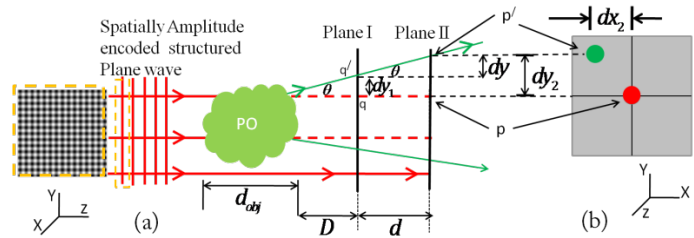

Fig.2. Schematic of the proposed technique. (a) Structured illumination is used to interrogate the PO and the traversed beam is intercepted at *Plane I* and *Plane II* . (b) *Plane II* (X-Y front), point $p$ (red dot) has moved to $p^/$ (green dot) when PO is introduced in the optical path

Here, the used structured illumination is a plane wave with spatially encoded *sinusoidal* amplitude as shown in Fig. 2. Two images are captured, simultaneously by using the proposed camera unit (see Section 3.1), focused at different planes, *Plane I* and *Plane II* respectively. Images captured on both the planes during the presence of the PO is compared (see Section 2.2.1) to the images of the same plane when there is no refractive gradient present in the optical path. By doing this the relative displacements of the structured pattern are estimated in both planes. For example, the point $p$ on *Plane II* gets displaced to $p^{/}$ when there is any distortion in the interrogating wavefront, see Figs. 2(a) and (b) and thus objective is to estimate the displacement vectors $[dx_2, dy_2]$. Similar observations are seen in *Plane I*, where displacement is from $q$ to $q^{/}$. The relative spatial displacement of the structured pattern on individual planes is estimated with the windowed Fourier transform (WFT) technique (see Section 2.2.1).

Now, plugging the Eq.(3) to Eq.(1), provides the formulation of phase estimation, which is

$$\nabla_{\perp}\varphi = \left[ k \int_{l} \nabla_{\perp} n(x,y) ds \right] = k \int_{l} \left( \frac{\partial n(x,y)}{\partial x} \hat{i} + \frac{\partial n(x,y)}{\partial y} \hat{j} \right) ds = k\mathbf{\theta} \quad .............................(5)$$

where $\mathbf{\theta} = \left[ \theta_x \hat{i} + \theta_y \hat{j} \right]$, beam deflection angles along x-axis $[\theta_x]$ and y-axis $[\theta_y]$ respectively. Now, the similar recipe of Section 2.1 can be used to solve the Poisson equation of Eq. 5 for estimating the phase, $\varphi$ of the deformed wavefront.

## 2.2.1. Estimation of relative displacement of the projected pattern and calculate beam deflection

The structured illumination has a spatially varying intensity profile that resembles a 2-dimensional sinusoidal crossed pattern as shown in Fig.2.2(a). The projected pattern at planes *Plane I* and *Plane II* along the optical axis gets distorted when it sees the PO in its path. The spatial distortion is calculated very accurately by exploiting the sinusoidal profile mathematically with the help of the windows Fourier transform technique [7, 19]. The displacement vectors $[dx_i, dy_i]$ of the distorted pattern are estimated space-continuously through the WFT technique, where the vertical and horizontal displacements are obtained from the horizontal and vertical strips of the recorded pattern. Two sets of displacement vectors, $[dx_1, dy_1]$ and $[dx_2, dy_2]$ are estimated from the images captured at *Plane I* and *Plane II* respectively, that are used to obtain the beam deflection angles $[\theta_x, \theta_y]$ through Eq.(4).

## 2.3 Estimation of density through iterative phase tomography

Phase tomography [18] and deflection tomography [33] schemes are widely used to reconstruct refractive index, n(r) [equivalently density, $\rho(r)$ in the ROI ] from measurements. In our case, the Phase tomography route is used to estimate density iteratively, from the measured phase $\varphi(x, y)$ obtained through the TIE, as mentioned in Sections 2.1 and 2.2. In which the following minimization problem is solved

Find $n(r)$, such that $\varphi_i = \frac{2\pi}{\lambda} \int_{0}^{L} n(\mathbf{r_i}) \, ds_i$; where $i = 1, 2, .......m$; number of rays ..................(6)

here, $\lambda$ is the wavelength and $L$ is the ray path length.

Eq.(6) represents the so-called Fermat's path that connects the source and detector, where the ray trajectory is governed by the Eikonal equation. A detailed explanation of phase tomography is provided in Ref. [18]. Recovered refractive profile, $n(r)$ is used in the Gladstone-Dale equation to estimate the density, $\rho(r)$ [18]. The details of various steps involved in density estimation are shown in the Flow-chart of Fig.3

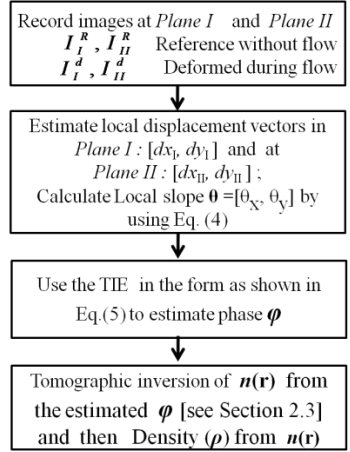

Fig.3: Flow chart shows the steps involved in density recovery

## 3. Experimental setup

This section consists of three major portions: first, the proposed camera optics that captures two images of the ROI simultaneously which are focused at different planes; secondly, the illumination optics that generates the structured light and finally, the experimental facility: the hypersonic shock tunnel, where the presented experiments are carried out.

### 3.1 Modified camera optics

Intensity images captured along the optical axis are the inputs for an intensity-based phase imaging scheme, where a complex optical arrangement containing several cameras are used to capture a dynamic event [22]. A much simpler approach that uses a single camera and customized split-beam arrangement (see Fig. 4) is introduced in this article that is capable of capturing two images of the object simultaneously that are focused at different planes. This arrangement suitably records experimental data that satisfies the need of the theoretical model proposed in Section 2. The schematic diagram of the proposed camera unit is shown in Fig.4 (a) and the actual unit in Fig. Fig.4 (b). The incident light wave collected by a primary lens ($f = 100\ mm$) is split into two streams with the help of a 50:50 beam splitter arrangement that direct the beams toward two plane mirrors M1 and M2 as shown in Fig 4(a). The reflected beams from M1 and M2 are further steered through the same beam splitter to the sensor plane (camera EOS M, Canon),  such that two non-overlapping images of the same object can be captured simultaneously, which are focused at different planes. The desired planes along the optical axis are focused by setting the positions of the mirrors M1 and M2, where the focusing plane at distance $u$ is associated with the image distance, $v$ through $u = \frac{fv}{v-f}$. Introducing a displacement, $\Delta v$ i.e. $v \pm \Delta v$ to the mirrors change the focusing plane from $u$ to $u_{new}\left(= u \mp \Delta u = \frac{f(v+\Delta v)}{v+\Delta v-f}\right)$

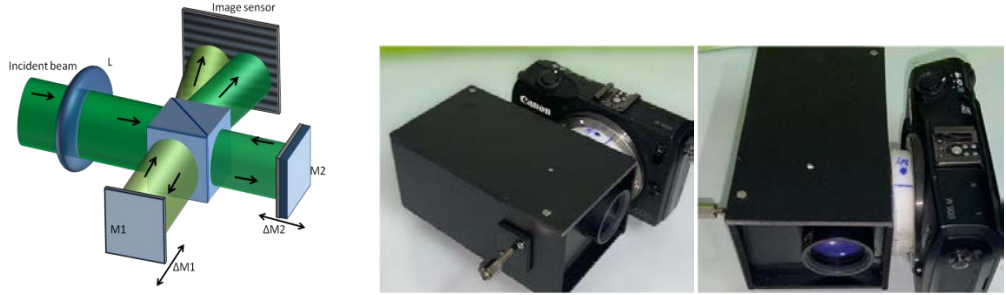

Fig. 4. Arrangement of optical component for the modular camera optics. (a) Schematic arrangement, M1, M2: mirrors, BS: beam splitter, L1: lens (b) Actual unit attached to the camera

The MCO fits the lens adapter of the camera unit and placed as shown in Figs.4 during experiments. Captured raw images of the ROI during flow diagnostics experiments are shown in Fig. 6. The captured images are aligned numerically by using the techniques mentioned in reference [34].

## 3.2 Structured illumination

The structured illumination, which is a plane wave that has spatially encoded amplitude is generated with the help of two-dimensional amplitude crossed-sinusoidal grating, light-emitting diode (LED), lens and parabolic mirror arrangements. The used arrangement is shown in Fig. 5. For the given setup, LED (~530nm±2nm) is used as a light source that is expanded with the help of lens L1, followed by the amplitude-grating (AG) and finally through the projection lens. The FNO of the projection lens is set to 10 equal to the FNO of the concave mirror used in experiments such that most of the emitted photons are utilised for visualization.

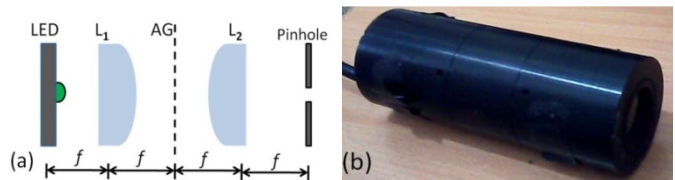

Fig.5. Structured illumination system, (a) schematic, (b) experimental module

## 3.3 Experiments in high-speed flow

High-speed flow experiments are carried out in the flow field generated in a hypersonic shock tunnel facility [7]. Hypervelocity flow of Mach no. M=8.5 is allowed to pass over an aerodynamic test model (hemispheric shape model of diameter 50mm) of interest placed in the test section, which is intercepted with the proposed optical technique. One side of the test section ( of size 300mm×300mm) is connected to the shock tube (50mm internal diameter) through the convergent-divergent nozzle and the other side to the dump tank. Before an operation, the driver and the driven side of the shock tube are separated with a metal diaphragm, that bursts open during experiments due to the applied pressure gradient across the diaphragm. Diaphragm bursting generates a primary shock wave that propagates towards the downstream of the driven section, where test gas is kept. The stagnation enthalpy (reservoir condition) of the test gas gets alleviated due to the interaction with the shock wave that further accelerates through the connected nozzle to the test section creating a flow field of the following parameters: $p_\infty$=341 Pa, $T_\infty$ =99.2 K, $M_\infty$=8.51 and enthalpy of H=1.54 MJ·kg−1. In the set of experiments presented in the paper, air and helium are used as the test- and driver gases respectively. The free-stream parameters of the flow field in the test section are tuned by changing the thickness of the primary diaphragm and so also by altering the gas used in experiments.

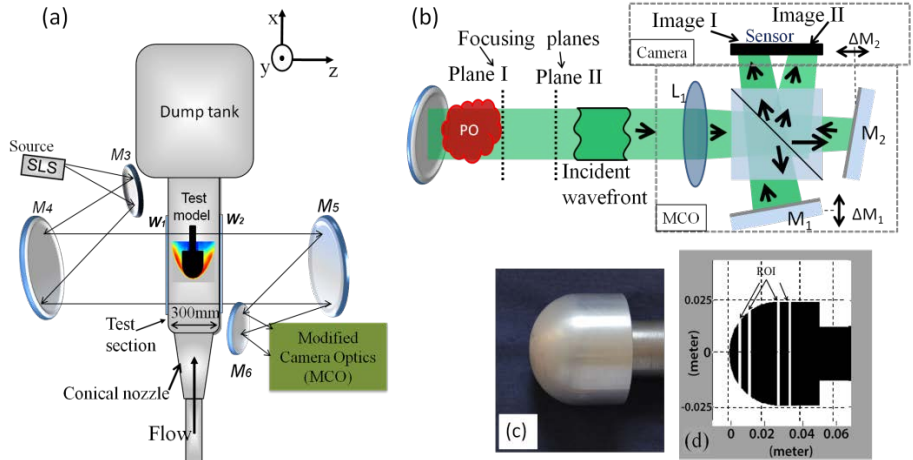

Fig.6. (a) Schematic of the experimental setup arranged around HST2 shock tunnel facility, (b) optical arrangement for capturing two images of flow field through the MCS, (c) Photograph of hemispheric test model (d) ROIs (5, 10, and 25 mm away from the tip of the model), across which the density profile is estimated

The interrogating beam coming out of lens L2 (Fig.5) of the structured light source (SLS) is collimated and directed by a concave mirror M4 to the test section to illuminate the ROI through the optical windows provided in the facility. Another concave mirror M5, kept on another side of the test section, collects the interrogating beam and direct it to the camera with the help of a plane mirror. All the optical components used in the experiment have a surface finish of $\lambda/10$. Two planes at z=100mm and z=1500mm from the centre of the test model are focused along the optical axis with the help of the camera optics. As mentioned earlier, the current setup helps in referencing the edge of the test model more accurately in comparison to BOS and shadow casting techniques as the ROI remain focused by one of the imaging optics of the proposed MCO, (z=100mm in this case). The other part of the imager is focused on a distant plane (z=1500mm) from the object that controls the sensitivity towards the refractive gradient of the imaging system. Two sets of images are captured in each plane, one with the flow and another as a reference without any disturbance within the optical path. The camera is triggered with the help of synchronization circuitry, designed for this purpose to capture the flow field at the steady-state region, with an exposure of 40us, sufficient enough to capture such flow fields [22]. Image pairs in each plane are processed for retrieving the local displacements ($[dx_i, dy_i]$), by following the technique of Section 2. Obtained displacement vectors from the planes are further processed to retrieve the slope of the deformed wavefront coming out of the flow field, which is processed further to retrieve the phase and finally the density distribution. Axis-symmetric assumption is used in the tomographic reconstruction. The details of the recovered density field are explained in the following section.

## 4. Experimental results and discussion

Images captured during the experiment are shown in Fig. 7, which are recorded by a single-sensor camera, into a single photograph using the dedicated camera optics. The two images are positioned on the sensor plane (hence in a single photograph) with minimal overlapping and their conjugate planes ( *Plane I* and *Plane II* ) are focused with the help of the focusing mirrors M1and M2 respectively. It is observed that these images are not registered properly and are found to be rotated and translated from the origin (Fig. 7(a)). Geometrical transformation is applied on the common ROIs of the images to bring them into a global coordinate system where they share the same x-y coordinates then cropped. The transformation [34] is attained through a calibration pattern, reference images in this case, by using the 'tie points' to align. These modified cropped images are shown in Fig. 7(b) and (c) for the hemispheric test model, which are the inputs to the algorithm presented in this paper.

Estimated density gradients retrieved for both the planes are shown in Fig.8, which resembles the schlieren images of the same flow field when captured with vertical and horizontal knife edges. The magnitude of displacement at *Plane II* is much higher than the *Plane I*, which is obvious as the in-plane displacement ($s$) of the projected pattern ($s = d * \tan\theta$) is more as it moves away from the flow field. The estimated 3D-density profile of the flow field is shown in Fig.10 and the cross-sectional plot of the same in Fig.9.

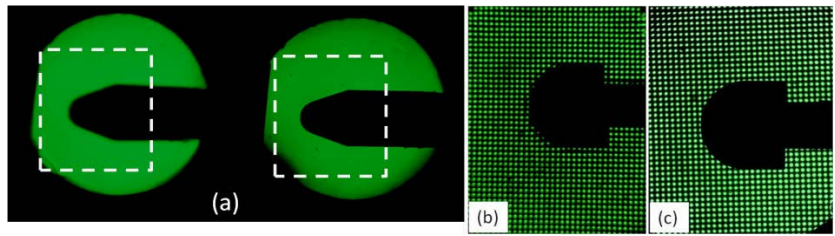

Fig.7. (a) Photograph of the conical model imaged through the MCO, where the planes *Plane I* and *Plane II* are focused by positioning the internal mirrors of the MCO unit. (b) and (c) represents the cropped images of the test object (see text) focused at *Plane I* and *Plane II* respectively, when structured illumination is on.

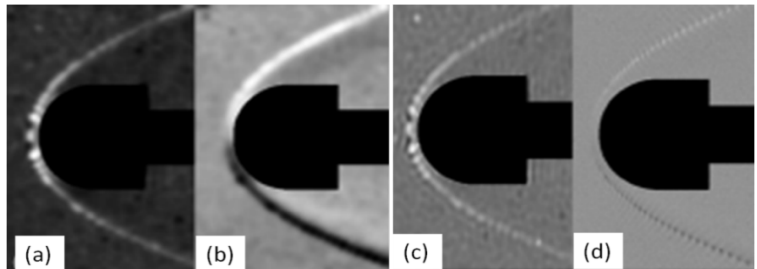

Fig.8. Estimated density gradients on the *Plane I* (a, b) and *Plane II* (c, d) respectively. Where, (a) and (c) represents - horizontal gradients; (b) and (d) are vertical gradients

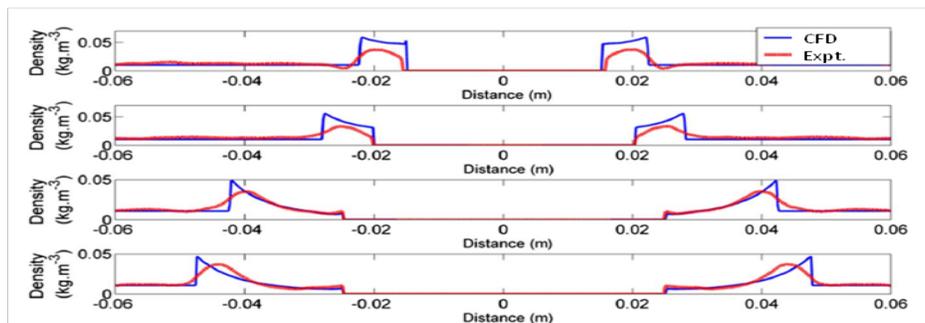

Fig.9: Estimated densities along with the ROIs of Fig. 6(d). A comparison is made with the CFD data for the same experiment.

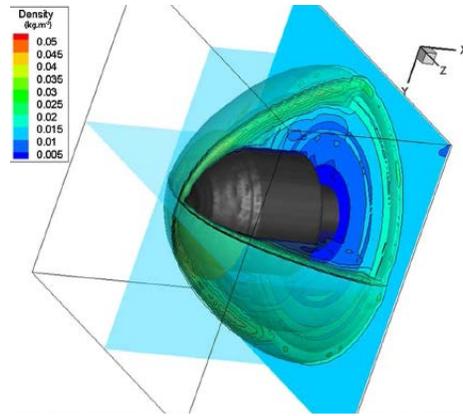

Fig.10. Recovered tomographic 3D density distribution of the flow field

The proposed method is further verified for accuracy and the experimentally recovered density is compared with the same flow field obtained through computational fluid dynamic (CFD) simulation. In this set of experiments, CFD generated data is considered to be the ground truth, where commercial software Fluent 13(R) is used for performing the CFD simulation [35]. Experimentally measured initial test conditions measured with the help of pressure and temperature transducers are used in setting the initial boundary conditions of the CFD simulation. In this case of simulation, the finite volume method is used while solving the conservation laws of fluids, which are Navier-Stokes equation for single component laminar viscous flow. No-slip adiabatic boundary conditions are incorporated at the fluid-body interface.

There are regions in the reconstructed density, especially the rapidly varying flow zones, the agreement with CFD is less. The probable reason for such less agreement is due to the finite exposure used during experiments or could be due to the smoothing process involved in the phase recovery. The overall reconstructed density of the flow field has a good agreement with the CFD simulated result, which confirms the utility of the proposed technique in a high-speed flow field.

The choice of using a plane wave with spatially encoded sinusoidal amplitude is backed up by our previous work [19], which also have mathematical support for processing sinusoidal pattern. However, any spatially varying amplitude pattern could be used for illumination as it merely helps to estimate local displacements of projection and also referencing information from the two planes. Perhaps, the speckle generated by a coherent source could be another choice.

## 5. Conclusions

In conclusion, we have demonstrated the utility of a structured light-based flow diagnostic tool for high-speed flow analysis. Flow-induced change in phase of the interrogating beam is estimated quantitatively through the spatial signature of the structured light probe, which is imaged in two planes by the modified camera optics. The recovered phase map from images is used in iterative phase tomography to recover three-dimensional density distribution that shows good agreement with the CFD simulated density for the same.

A new technique is presented to obtain the intensity derivative for TIE calculation through the use of the spatial signature of the structured light beam. In the proposed technique, the need for small defocused distance to estimate FD approximated intensity derivative, $\partial I/\partial z$ is alleviated, since the $\partial I/\partial z$ is now estimated through the spatial variations of the geometrical projection of structured light, imaged in two planes, and thus larger separation distance $(\partial z)$ is admissible. Experiments conducted in the HST2 facility confirms the claim. The proposed technique for phase estimation (hence density) is suitable for an imaging system that has a larger $f/\#$, as in the case of flow visualization. Moreover, the construction of the technique while implementing in flow visualization keeps the ROI focused, which is imaged by *Plane I* in this case, helps in referencing the test model accurately. Thus we believe, the proposed technique is useful for diagnosing high-speed flows and also worth investigating for other applications in future.

**Acknowledgement**: The presented research work was carried out by the corresponding author in the year 2018 during the post-PhD period under the supervision of Prof. K.P.J. Reddy in the Department of Aerospace Engineering, Indian Institute of Science, Bangalore.